\documentclass[twocolumn,pra]{revtex4}
\usepackage{graphicx}
\usepackage{amsmath}

\begin{document}
\title{A New Class of  non-Hermitian Quantum Hamiltonians with ${\cal PT}$ symmetry} \author{Katherine Jones-Smith}
\author{Harsh Mathur}
\affiliation{Department of Physics, Case Western Reserve University,10900 Euclid Avenue, Cleveland OH 44106-7079}

\begin{abstract}
We formulate quantum mechanics for non-Hermitian Hamiltonians that are	
invariant under	${\cal PT}$, where ${\cal P}$ is	the parity and ${\cal T}$
denotes time reversal, for the case that time reversal symmetry is odd	
(${\cal T}^2 = -1$), generalizing prior work for the even case
(${\cal T}^2 = 1$). We discover	an analogue of Kramer's theorem	
for ${\cal PT}$	quantum	mechanics, present a prototypical example
of a ${\cal PT}$ quantum system with odd time reversal, and discuss
potential applications of the formalism.
\end{abstract}

\maketitle

The basic structure of quantum mechanics was delineated in the
early days of the theory \cite{dirac} and it has not been modified since. Still
it is desirable to ask whether the structure can
be altered and generalized. For example %linearity and 
%superposition are core principles of quantum mechanics. Nonetheless 
Weinberg showed that it is possible to formulate
a non-linear generalization of quantum mechanics and to thereby
subject the linearity of quantum mechanics to a quantitative test \cite{weinberg}.
A fruitful generalization of the canonical principles, 
was the discovery that particles can have
fractional statistics that interpolate between Bose and Fermi, albeit only in
two spatial dimensions \cite{wilczek}.
More recently the principle that the Hamiltonian and other observables should
be represented by Hermitian operators has been re-examined \cite{bender}. 
A non-Hermitian formulation of quantum mechanics promises to 
%, if it were possible, would 
enlarge the set of possible Hamiltonians that
physicists could deploy to describe fundamental physics beyond the
standard model or for the effective description of condensed matter phenomena.
%of complex many-body
%phenomena in condensed matter physics.

Bender and co-workers have shown that the assumption of
hermiticity can be relaxed under the circumstance that the Hamiltonian
is invariant under the combined symmetry ${\cal PT}$ where ${\cal P}$
denotes the parity and ${\cal T}$ denotes time reversal provided that in
addition the eigenvalues of the Hamiltonian are real and the appropriately
defined left and right eigenvectors coincide. By now many 
examples of ${\cal PT}$ quantum mechanics have been presented in 
the literature \cite{bender}. 

In all work on ${\cal PT}$ quantum mechanics
to date it has been implicitly assumed that time reversal is even, 
${\cal T}^2 = 1$. However in quantum theory this is only true of bosonic
systems with integer spin. For fermionic systems, with half-integer spin, 
time-reversal is odd, ${\cal T}^2 = -1$ \footnote{The proposition that there are two 
kinds of time-reversal is proved as follows: We assume that time-reversal is anti-linear.
Thus ${\cal T} \psi = L \psi^{\ast}$ where $L$ is a linear operator. Next assume that ${\cal T}^2$ leaves
all states unchanged up to a phase $\exp(i \phi)$. It follows $L L^{\ast} = \exp(i \phi)$. This implies
$L = \exp(i \phi) L^{-1}$. By conjugation and using $(L^{-1})^{\ast} = (L^{\ast})^{-1}$ it follows that
$L L^{\ast} = \exp ( - i \phi)$ also. Thus $\exp(i \phi) = \pm 1$ as claimed. In textbook hermitian 
quantum mechanics it is assumed that $L$ is unitary. The proof given here does not make that 
assumption and hence generalizes the proposition to the non-hermitian case.}. Quarks and leptons
in particle physics, approximately half of all nuclei and atoms, and a plethora
of condensed matter problems including magnetic spin models and
solid state electronic matter fall into this 
category. Thus it is clearly important
to generalize the construction of ${\cal PT}$ quantum mechanics to the 
case that ${\cal T}$ is odd. That is the purpose of this Communication.

Before embarking on the generalization it is useful to recall the key
principles of ${\cal PT}$ quantum mechanics for the case that time 
reversal is even. For simplicity let us assume that the Hilbert space
of states has finite dimension $N$ so that the state of the system may
be specified by a wavefunction that is an $N$ component column vector
with complex components $\psi (n)$, with $n = 1, 2, 3, \ldots N$. We will work
in a basis such that the operation of time reversal consists of simply
taking the complex conjugate of the wavefunction, ${\cal T} \psi = 
\psi^{\ast}$; thus ${\cal T}^2 = 1$.
It can be shown that such a basis can always be found.
Since parity is a linear operator it may be represented by a matrix that we
shall denote $S$; thus ${\cal P} \psi = S \psi$. We assume that parity commutes
with ${\cal T}$ and that parity applied twice is the identity transformation. 
It follows from these
considerations that the matrix $S$ is real and that its square is 
the identity. Without loss of generality we may work in a basis where the
parity matrix $S$ is diagonal. Since $S^2 = 1$, in this basis the matrix $S$ 
will have diagonal entries $\pm 1$. 
%\footnote{That a basis can be found in which
%$S$ is diagonal and ${\cal T}$ consists of conjugation follows if we make
%the additional assumption that $S$ is unitary. Then it follows $S$ is not only real
%but also symmetric and can therefore be diagonalized 
%by an orthogonal transformation. Furthermore one can also show that the form
%of time reversal remains simple complex conjugation if the change of basis 
%is an orthogonal transformation \cite{messiah}.}. 
For definiteness let us assume
that $N$ is even and ${\cal P}$ has the form
\begin{equation}
S = \left(
\begin{array}{cc}
{\cal I} & 0 \\
0 & -{\cal I}
\end{array}
\right)
\label{eq:peven}
\end{equation}
where ${\cal I}$ denotes the $N/2$ dimensional identity matrix.

Now in quantum mechanics one conventionally
defines the inner product of two states as
$( \phi, \psi ) = \phi^{\dagger} \psi = \sum_{n=1}^{N} \phi^{\ast} (n) \psi (n)$.
However in ${\cal PT}$ quantum mechanics a different inner product, one
that is determined by the Hamiltonian, is physically relevant. 
As a prelude to the formulation of 
this inner product it is useful to first introduce the $PT$ inner
product defined as $(\phi, \psi)_{PT} \equiv ({\cal PT} 
\phi)^{T} \psi = \phi^{\dagger} S \psi$.
From this definition it is easy to verify that $(\phi, \phi)_{PT}$ is real but with
indefinite sign. Thus the $PT$ inner product is not by itself a viable inner product
to use in quantum mechanics. 

Conventionally one requires the Hamiltonian $H$ (and all other observables
as well) to be hermitian, $H^{\dagger} = H$. The principal new feature of ${\cal PT}$ 
quantum mechanics is that it is acceptable for the Hamiltonian to be non-Hermitian
provided it satisfies three additional criteria instead. First, the Hamiltonian must
be invariant under ${\cal PT}$ ({\em i.e.} it must commute with ${\cal PT}$). This
enforces the following block structure on the Hamiltonian matrix
\begin{equation}
H = \left(
\begin{array}{cc} 
A & i B \\
i C & D
\end{array}
\right)
\label{eq:pthamiltonian}
\end{equation}
where $A, B, C$ and $D$ are real matrices. It follows from invariance under
${\cal PT}$ that the eigenvalues of $H$ come in conjugate  pairs for if $\phi$ is
an eigenfunction with eigenvalue $\lambda$, then ${\cal PT} \phi$ is an 
eigenfunction with eigenvalue $\lambda^{\ast}$.
The second criterion is that ${\cal PT}$ must be
unbroken in the sense that it should be possible to find eigenvectors of the
Hamiltonian, $\psi_i$, that are invariant under ${\cal PT}$ ({\em i.e.} 
${\cal P} {\cal T} \psi_i = \psi_i$). Evidently if ${\cal PT}$ is unbroken then
the eigenvalues must all be real. Conversely, if the eigenvalues are real
one can show that ${\cal PT}$ is unbroken. Note that a state that is
invariant under ${\cal PT}$ will have the form
\begin{equation}
\psi = \left( \begin{array}{c}
\xi \\
i \eta
\end{array} \right)
\label{eq:invariantvector}
\end{equation}
where $\xi$ and $\eta$ are real column vectors with $N/2$ components
each and we are assuming that $S$ has the form given in eq (\ref{eq:peven}).
Proving that ${\cal PT}$ is unbroken (or equivalently that the eigenvalues are
real) is frequently the most difficult step in ${\cal PT}$ quantum 
mechanics \footnote{Hermiticity is a sufficient condition for the eigenvalues
to be real but not necessary. Nonetheless, 
a generic non-Hermitian operator has complex eigenvalues; see, 
for example, M.L. Mehta, {\em Random Matrix Theory} (Academic Press,
New York, 1991).}.
The third criterion stems from the observation that $H$ and $H^T$ have the
same eigenvalues and if $H$ is invariant under ${\cal PT}$ then the eigenvectors
$\psi$ of $H$ and $\phi$ of $H^T$ are orthogonal under the $PT$ inner
product unless they correspond to conjugate eigenvalues
\footnote{This observation is similar to, but 
distinct from, the familiar biorthogonality theorem that for a
non-hermitian matrix $H$, the eigenvectors of $H$ and $H^{\dagger}$ 
are orthogonal to each other under the standard inner product unless
they correspond to conjugate eigenvalues. See P.M. Morse and H. Feshbach, 
{\em Methods of Mathematical Physics} (McGraw-Hill, New York, 1953) for 
a discussion of the conventional bi-orthogonality theorem.}. Thus $H$ and 
$H^T$ are said to be ``dual'' under the $PT$ inner product and the third
condition imposed on the Hamiltonian in $PT$ quantum mechanics is 
that the Hamiltonian should be self dual {\em i.e.} $H = H^T$. This ensures
that the eigenvectors of $H$ and $H^T$ are the same and are orthogonal
to each other under the ${\cal PT}$ inner product.

Having specified the kinds of Hamiltonians that are permissible in 
${\cal PT}$ quantum mechanics we now return to the formulation 
of the appropriate inner product. Eigenvectors of $H$ will fall into
two classes under the $PT$ inner product; namely, those for which
$(\psi_i, \psi_i)_{PT}$ is positive and those for which $(\psi_i, \psi_i)_{PT}$ 
is negative \footnote{It is also possible that there are eigenvectors that
are orthogonal to themselves under the $PT$ inner product. In the absence of
degeneracies such an orthogonality is a catastrophe in the sense that
it is then impossible to formulate ${\cal PT}$ quantum mechanics for 
the Hamiltonian under consideration.}. We now introduce a linear operator
denoted ${\cal C}$ which has the property ${\cal C} \psi_i = s_i \psi_i$ where
$\psi_i$ are eigenvectors of $H$ and $s_i$ is the sign of the $PT$ norm
of the eigenvector, $(\psi_i, \psi_i)_{PT}$. We have defined ${\cal C}$ by its action on the
eigenvectors of $H$ but since it is a linear operator it must be represented
by some matrix in the standard basis; we denote this matrix $K$ ({\em i.e.}
${\cal C} \psi = K \psi$). Evidently the operator ${\cal C}$ 
commutes with the combination ${\cal PT}$ although it may not 
commute with either ${\cal P}$ or ${\cal T}$ separately. In terms of matrices
this corresponds to the statement $K S = S K^{\ast}$.  
Furthermore ${\cal C}^2 = 1$. It is evident from the definition given
here that the Hamiltonian plays a crucial role in determining the operator
${\cal C}$.

We now define the $CPT$ inner product
$(\phi, \psi)_{CPT} \equiv ({\cal CPT} \phi)^T \psi = \phi^{\dagger} K^T S \psi$.
This is the inner product used in ${\cal PT}$ quantum mechanics in lieu of the
standard inner product. 
Evidently all states have a positive norm with this inner product; this is most
easily seen by expanding in the eigenbasis of the Hamiltonian. Furthermore
the $CPT$ norm of any state is preserved under time evolution via Schr\"{o}dinger's
equation $i \partial \psi/\partial t = H \psi$. Thus it is possible to consistently
formulate quantum mechanics using the $CPT$ inner product, notwithstanding
the non-hermiticity of the Hamiltonian.

Finally we note that in quantum mechanics it is customary to stipulate that
observables besides the Hamiltonian should also be represented by hermitian
operators. In ${\cal PT}$ quantum mechanics it is no longer required that
observables be hermitian. Rather we define the $CPT$ adjoint $A^{\star}$ 
of an operator $A$ by imposing the
 condition that $(\phi, A \psi)_{CPT} = 
(A^{\star} \phi, \psi)_{CPT}$ for all states $\phi$ and $\psi$. Observables 
are then required to be $CPT$ self-adjoint, $A = A^{\star}$. This is sufficient
to ensure that the eigenvalues of $A$ are real and that the usual principles
of quantum measurement and uncertainty relations \cite{kate}
may be applied even though the observables are no longer hermitian 
in the usual sense.

This concludes our resume of the principles of ${\cal PT}$ quantum mechanics
for the case of even time reversal symmetry. We now enunciate the corresponding
principles for the case that time reversal symmetry is odd.

For simplicity we assume that the Hilbert space of states has finite dimension $2N$ so
that the state of the system may be specified by a wavefunction that is a $2N$ 
component column vector with complex components $\psi(n)$ with
$n = 1, 2, 3, \ldots 2N$. We will work in a basis such that the operation of 
time reversal is given by
\begin{equation}
{\cal T} \psi = Z \psi^{\ast}
\label{eq:todd}
\end{equation}
In the following it will be convenient to refer to $2 \times 2$ matrices as quaternions.
Any quaternion can be expanded as $q = q_0 \sigma_0 + i q_1 \sigma_1 + 
i q_2 \sigma_2 + i q_3 \sigma_3$ where $\sigma_0$ is the $2 \times 2$ identity
matrix and $\sigma_1, \sigma_2$ and $\sigma_3$ are the Pauli matrices. 
The coefficients $q_0, q_1, q_2$ and $q_3$ are complex in general. In the
case that they are real, the quaternion is said to be real. By suitably partitioning
a $2 N \times 2 N$ matrix into $2 \times 2$ blocks one can view it as an $N \times N$
matrix of quaternions. The matrix $Z$ for example is an $N \times N$ 
quaternion matrix with every diagonal term equal to $i \sigma_2$ and every
off-diagonal term equal to zero. It can be shown that a basis can always be found
in which time reversal has the canonical form eq (\ref{eq:todd}).
Note that in this case ${\cal T}^2 = -1$. 

As in the even case we assume that
parity is a linear operator represented by a matrix $S$. We again assume that 
parity commutes with time reversal and that parity applied twice is the identity
transformation. It follows from these considerations that the matrix $S$ is 
quaternion real and that its square is the identity. Without loss of generality
we may work in a basis where simultaneously time reversal has the canonical form 
eq (\ref{eq:todd}) and the parity matrix $S$ is diagonal. In this basis the matrix
$S$ will have diagonal entries $\pm 1$. These entries come in pairs for if 
$\psi$ is an eigenvector of $S$ then so is ${\cal T}$. This follows by virtue of
the quaternion reality of $S$ and is an example of Kramers theorem in 
quantum mechanics. Thus the number of positive parity states as well as
negative parity states must be even. For definiteness we will assume that
$N$ is even and that $S$ has the form eq (\ref{eq:peven}) except that
${\cal I}$ now represents an $N \times N$ identity matrix. 

Having described time-reversal and parity our next task is to generalize
the definition of $PT$ inner product to the case that time reversal is odd.
We define $(\phi, \psi)_{PT} \equiv ({\cal PT} \phi)^T Z \psi = \phi^{\dagger} S \psi$;
note the crucial insertion of a factor of $Z$ in the definition. With this definition
it follows that the states $\psi$ and ${\cal PT} \psi$ are orthogonal to each
other and have the same norm. The norm of any state $( \phi, \phi)_{PT}$ is
real but of indefinite sign. Thus, as in the even case, the $PT$ inner product
is not by itself a viable inner product to use in quantum mechanics but is
merely a stepping stone towards a viable inner product.

It is noteworthy that the state $\phi$ and ${\cal PT} \phi$ are also orthogonal
to each other under the standard inner product. This shows that the two
states are truly distinct and linearly independent. Furthermore one can
show that it is impossible to find a state that is invariant under ${\cal PT}$
in the case of odd time reversal. However one can assemble the two 
column vectors $\phi$ and $- {\cal PT} \phi$ into a ``$PT$ doublet'',
$ ( \phi, - {\cal PT} \phi)$. The doublet is a $2 N \times 2$ matrix that
can equivalently be viewed as a single $N$ component column of
quaternions. It is easy to see that the $PT$ doublet has the form 
shown in eq (\ref{eq:invariantvector}) where $\xi$ and $\eta$ now 
both represent columns of real quaternions. Under the application of ${\cal PT}$,
the $PT$ doublet gets post-multiplied by a single
Pauli matrix $i \sigma_2$. The $PT$ doublet is thus the closest
analogue to an invariant state in ${\cal T}$-even ${\cal PT}$ 
quantum mechanics. 

We now explore the conditions that must be imposed on a 
Hamiltonian for it to be viable in odd ${\cal PT}$ quantum mechanics.
First of course it must be invariant under ${\cal PT}$ ({\em i.e.} it must 
commute with ${\cal PT}$). This forces the 
Hamiltonian to have the form eq (\ref{eq:pthamiltonian}) but with $A, B, C$
and $D$ now $N \times N$ matrices composed of real quaternions.
It follows from invariance that the eigenvalues of $H$ come in conjugate
pairs for if $\phi$ is an eigenfunction with eigenvalue $\lambda$, then
${\cal PT} \phi$ is an eigenfunction with eigenvalue $\lambda^{\ast}$.
The second condition is that ${\cal PT}$ must be unbroken which in this
case means that the eigenstate $\phi$ and ${\cal PT} \phi$ must be degenerate.
Evidently if ${\cal PT}$ is unbroken the eigenvalues must be real. Conversely
if the eigenvalues are real it follows that ${\cal PT}$ is unbroken, a statement
that is the ${\cal PT}$ quantum analogue of Kramer's theorem in standard
quantum mechanics. The final condition stems from the observations
that (i) $H$ and $Z H^T Z^T$ have the same eigenvalues; and (ii) if $H$ is invariant
under ${\cal PT}$, then the eigenvectors $\psi$ of $H$ and $\phi$ of $Z H^T Z^T$
are orthogonal under the $PT$ innerproduct, unless they correspond to conjugate
eigenvalues. Thus $H$ and $ZH^T Z^T$ are said to be ``dual'' under the 
$PT$ inner product and the third condition imposed on the Hamiltonian in 
$PT$ quantum mechanics is that the Hamiltonian should be self dual, {\em i.e.}, 
$H = Z H^T Z^T$. This ensures that the eigenvectors of $H$ and $Z H^T Z^T$ are
the same and are appropriately 
orthogonal to each other under the ${\cal PT}$ inner product.

Having identified the kinds of Hamiltonians that are permitted in 
${\cal PT}$ quantum mechanics for odd time reversal we now formulate
the appropriate inner product. $PT$ doublets $(\phi, -{\cal PT}\phi)$
fall into two classes under the
$PT$ inner product: those for which $(\phi_i, \phi_i)_{PT}$ is positive and
those for which it is negative. As in the even case we introduce a new
linear operator ${\cal C}$ with a corresponding matrix $K$ which has the
defining property that ${\cal C} \psi_i = s_i \psi_i$ where $\psi_i$ denotes
an eigenvector of $H$ and $s_i$ is the sign of the $PT$ norm of that 
eigenvector, $(\psi_i, \psi_i)$. ${\cal C}$ commutes with ${\cal PT}$ 
corresponding to the statement $K S Z = S Z K^{\ast}$. As in the even
case ${\cal C}^2 = 1$. 

We now define the $CPT$ inner product for the odd case as 
$(\phi, \psi)_{CPT} \equiv ({\cal CPT} \phi)^T Z \psi = \phi^{\dagger} K^T S \psi$. 
As in the even case, all states have a positive norm with this inner product; moreover
the norm of any state is preserved under Schr\"{o}dinger time evolution.
Thus it is possible to consistently formulate quantum mechanics using the
$CPT$ inner product as defined here for the case of systems that are odd
under time reversal. 

Finally our remarks about operators corresponding to other observables 
besides the Hamiltonian for the case of even time reversal may be applied 
mutatis mutandis to the case of odd time reversal. This concludes our
formulation of ${\cal PT}$ quantum mechanics for the odd case. 
Although for simplicity we have only discussed finite dimensional
Hilbert spaces, the extension to the infinite dimensional case is 
straightforward.

To illustrate these principles it is helpful to consider the simplest
non-trivial examples of ${\cal PT}$ quantum mechanics corresponding
to $N = 2$ for the even case and $N=4$ for the odd case; the two-level
model for the even case has been discussed before in ref \cite{berry}.
For the even case the most general Hamiltonian matrix that meets all
the conditions of ${\cal PT}$ quantum mechanics is
\begin{equation}
H = \left(
\begin{array}{cc}
a & i b \\
i b & -a
\end{array}
\right)
\label{eq:twolevel}
\end{equation}
Here $a$ and $b$ are real numbers and 
we have imposed the additional condition that $H$ is traceless
for simplicity \footnote{If $H$ has a trace it can always be written as the
trace times the identity plus a traceless part. Note that the trace term does
not affect the eigenvectors and shifts all the eigenvalues by a constant value.
Thus the effects of the trace can be trivially incorporated.}. 
Note that for $b \neq 0$ this matrix is 
explicitly non-hermitian. It is instructive to compare eq (\ref{eq:twolevel}) to 
the most general two-level hermitian Hamiltonian that is invariant under
even time reversal. That Hamiltonian is obtained from eq (\ref{eq:twolevel}) by
replacing the pure imaginary off diagonal terms with pure real ones ({\em i.e.}
$ i b \rightarrow b$). 
The eigenvalues of $H$ are $\pm \sqrt{ a^2 - b^2}$.
Thus $PT$ is unbroken only for $a^2 > b^2$. So long as this
condition is satisfied the Hamiltonian $H$ may be parametrized
as $a =\rho \cosh(\chi)$ and $b = \rho \sinh \chi$ where $\rho > 0$
and $-\infty < \chi < \infty$. This parametrization applies for $a > 0$ which
we will assume hereafter. The case $a < 0$ can be parametrized and analyzed
in exactly the same way. The eigen-matrix is 
\begin{equation}
U = \left(
\begin{array}{cc}
\cosh \chi/2 & \sinh \chi/2 \\
i \sinh \chi/2 & i \cosh \chi/2
\end{array}
\right)
\label{eq:eveneigenmatrix}
\end{equation}
Here the first column corresponds to the eigenvector with positive eigenvalue
$\rho$ and the second to the negative eigenvalue $-\rho$; note that the eigenvectors
have the $PT$ invariant form in eq (\ref{eq:invariantvector}). It is easy to verify
that the positive eigenvector also has positive $PT$ norm; the negative has 
negative norm. Thus the operator ${\cal C}$ is simply the normalized Hamiltonian
({\em i.e.} $H$ divided by the magnitude of the eigenvalues $\sqrt{a^2 - b^2}$),
\begin{equation}
{\cal C} = \left(
\begin{array}{cc}
\cosh \chi & i \sinh \chi \\
i \sinh \chi & - \cosh \chi
\end{array}
\right)
\label{eq:evenc}
\end{equation}
Finally the most general operator $A$ that corresponds to an observable by
virtue of being $CPT$ self-ajoint is 
\begin{equation}
A = \left( \begin{array}{cc}
A_0 + A_3 - i A_1 \tanh \chi & A_1 - i A_2 + i A_3 \tanh \chi \\
A_1 + i A_2 + i A_3 \tanh \chi & A_0 - A_3 + i A_1 \tanh \chi
\end{array}
\right)
\label{eq:evenobservable}
\end{equation}
Note that in the limit $\chi \rightarrow 0$, the most general observable is 
simply a hermitian matrix; in the same limit the Hamiltonian $H$ becomes
hermitian as well. 

Finally let us consider the simplest non-trivial example of ${\cal PT}$ 
quantum mechanics for the case of odd time reversal symmetry with 
$N = 4$.
The most general traceless Hamiltonian matrix that meets the criteria
of being invariant and self dual under ${\cal PT}$ is given by
\begin{equation}
H = \left(
\begin{array}{cc}
a & i b \\
i b^{\dagger} & -a
\end{array}
\right)
\label{eq:fourlevel}
\end{equation}
where $b = b_0 \sigma_0 + i b_1 \sigma_1 + i b_2 \sigma_2 + i b_3 \sigma_3$ is 
a real quaternion, 
and $a = a_0 \sigma_0$ is a real quaternion proportional to the identity.
It is instructive to compare this Hamiltonian to the most general four-level
hermitian Hamiltonian that is invariant under odd time-reversal; the latter 
is obtained by replacing the pure imaginary off diagonal quaternions with
pure real quaternions, $i b, i b^{\dagger} \rightarrow b, b^{\dagger}$.
It is also instructive to compare eq (\ref{eq:fourlevel}) to its counterpart in 
the even case eq (\ref{eq:twolevel}) obtained by replacing the real quaternions
$a$ and $b$ with real numbers.  
%The most general traceless Hamiltonian matrix that meets the criteria
%of being invariant and self dual under ${\cal PT}$ is still given by
%eq (\ref{eq:twolevel}) provided we interpret $b$ as a real quaternion,
%$b_0 \sigma_0 + i b_1 \sigma_1 + i b_2 \sigma_2 + i b_3 \sigma_3$, 
%and $a$ as a real quaternion proportional to the identity, $a = a_0 \sigma_0$. 
%It is instructive to compare this Hamiltonian to the most general four-level
%hermitian Hamiltonian that is invariant under odd time-reversal; the latter 
%is obtained by replacing the pure imaginary off diagonal quaternions with
%pure real quaternions, $i b \rightarrow b, b^{\dagger}$. 
The eigenvalues of the ${\cal PT}$
invariant Hamiltonian are $\pm \sqrt{ a^2 - b^2 }$ where $a^2 = a_0^2$ and
$b^2 = b_0^2 + b_1^2 + b_2^2 + b_3^2$ denote the magnitudes of the
quaternions $a$ and $b$. Thus $PT$ is unbroken only for $a^2 > b^2$.
So long as this condition is met (and $a_0 > 0$; the case $a_0 < 0$ can 
be analysed similarly) we can parametrize the ${\cal PT}$ Hamiltonian
by writing $a_0 = \cosh \chi$ and adopting polar co-ordinates
$(\sinh \chi, \varphi, \theta, \phi)$ in the four dimensional space of
the components of $b$ so that 
$b_0 = \sinh \chi \cos \varphi, 
b_3 = \sinh \chi \sin \varphi \cos \theta, 
b_1 = \sinh \chi \sin \varphi \sin \theta \cos \phi$ and 
$b_2 = \sinh \chi \sin \varphi \sin \theta \sin \phi$. 
In terms of this parametrization the eigenmatrix has the form
\begin{equation}
U = \left( 
\begin{array}{cccc}
q \cosh \chi/2 & q \sinh \chi/2 \\
i q p \sinh \chi/2 & i q p \cosh \chi/2
\end{array}
\right)
\label{eq:eigenmatrixodd}
\end{equation}
Here $q$ is the real quaternion corresponding to a rotation about
the $n_x = \sin \phi, n_y = - \cos \phi, n_z = 0$ axis by an angle of
$\theta$; and $p = \exp(-i \varphi \sigma_z)$, a rotation about the $z$ -axis
by an angle $2 \varphi$. 
The first two columns correspond to the positive energy ${\cal PT}$ 
doublet; the second two to the negative energy doublet. It is easy to 
verify that the positive doublet also has positive $PT$ norm; the negative
has negative norm. Thus the operator ${\cal C}$ coincides with the
normalized Hamiltonian ({\em i.e.} $H$ divided by $\sqrt{a^2 - b^2}$).
Finally, the most general operator B that corresponds to an 
observable by virtue of being $CPT$ self-adjoint is
\begin{equation}
B = \left( \begin{array}{cc}
q & 0 \\
0 & q p
\end{array} 
\right) A \left( \begin{array}{cc}
q^{\dagger} & 0 \\
0 & p^{\dagger} q^{\dagger} 
\end{array}
\right)
\label{eq:oddobservable}
\end{equation}
where $A$ is still given by eq (\ref{eq:evenobservable}) but with $A_0, A_1, A_2$ and
$A_3$ now interpreted as arbitrary $2 \times 2$ hermitian matrices.

It is worth recalling that in conventional quantum mechanics a variety
of complicated quantum mechanical problems can be truncated to a 
two level model \cite{feynman}. Thus the two and four level models
presented here should be regarded not merely as toy models but as
effective Hamiltonians that can be used as the basis for further investigation
of the quantum dynamics of ${\cal PT}$ quantum systems. 

In summary we have generalized the construction of ${\cal PT}$ quantum
mechanics to the case that time-reversal symmetry is odd. We hope this
generalization will further stimulate the search for natural phenomena that are
described by ${\cal PT}$ quantum mechanics. The most important
model in fundamental physics that is odd under time-reversal is the Dirac
equation. It is natural to ask whether the formulation of ${\cal PT}$
quantum mechanics presented here may lead to a new form of Dirac's equation.
The results of that investigation will be reported elsewhere \cite{ptachyon}.

We acknowledge a stimulating conversation with Carl Bender.

\end{document}